# Coherent perfect absorption and laser modes in a cylindrical structure of conjugate metamaterials

Yangyang Fu[1, 2, 3], YadongXu[1], Huanyang Chen[2] and Steven A. Cummer[3,]

*1. College of Physics, Optoelectronics and Energy, Soochow University, No.1 Shizi Street, Suzhou 215006, China.*

*2. Institute of Electromagnetics and Acoustics and Department of Electronic Science, Xiamen University, Xiamen 361005, China.*

*3. Department of Electrical and Computer Engineering, Duke University, Durham, North Carolina 27708, USA*

E-mail: *ydxu@suda.edu.cn* or *kenyon@xmu.edu.cn* or *cummer@ee.duke.edu*

**Abstract**

In this work, we theoretically find that coherent perfect absorption (CPA) and laser modes can be realized in a two-dimensional cylindrical structure composed of conjugate metamaterials (CMs). The required phase factors of CMs for achieving CPA and laser modes are determined by the geometric size of the CM cylinder, which is a unique feature compared with other non-Hermitian optical systems. Based on this property, we also demonstrate that CPA and laser modes can exist simultaneously in a CM cylinder with an extremely large size, where the excitations of CPA and laser modes depend on the angular momentum of coherent incident light. Therefore, compared with the well known parity time symmetry, our work opens up a brand-new path to obtaining CPA and laser modes, and is a significant advance in non-Hermitian optical systems.

## 1. Introduction

In an optical cavity system with gain media [1], when light amplification reaches a threshold value (i.e., when laser modes are excited), coherent electromagnetic (EM) radiation can be emitted from the cavity. In contrast, as the time-reversed counterpart of laser modes, coherent perfect absorption (CPA) [2, 3] (fully absorbing incoming coherent waves) was proposed by reversing the gain media with absorption (loss) one in the same optical cavity system. Furthermore, by applying the concept of parity-time (PT) symmetry into optics [4-10], CPA and laser modes [11, 12] have been realized in non-Hermitian optical systems, where the modulated loss and gain are balanced ($n(-r) = n(r)^*$) in artificial materials.

Such PT-symmetric CPA and laser modes were experimentally demonstrated in a single cavity [13]. CPA has been widely investigated in many domains, such as thin films [14, 15], metasurfaces [16], and surface plasmons [17]. In particular, the concept of CPA has been extended to two-dimensional cylindrical structures [18-20], where incident waves with angular momenta can be totally absorbed. However, to the best of our knowledge, CPA and laser modes, especially for obtaining CPA and laser modes simultaneously, have not been investigated in two-dimensional cylindrical structures. Therefore, a new mechanism to realize CPA and laser modes in two-dimensional scale is worth seeking.

On the other hand, metamaterials (including planar metasurfaces), composed of engineered units in subwavelength scale, allow the control of wave dynamics in optics [21-23], yet often with intrinsic losses. Nevertheless, we can flexibly construct non-Hermitian metamaterials with gain or loss, hence with new interesting optical phenomena, such as gain-loss induced stable surface waves [24]. Recently conjugate metamaterials (CMs) ($\varepsilon(x) = \mu^*(x)$) [25], as a kind of non-Hermitian metamaterials [26], which involve optical loss and gain simultaneously but possess purely real refractive indexes [27], have been studied extensively [28-31]. Such media exhibit special optical features. For example, they are either positive index media or negative index media, depending on complex angles (or phase factors) of the material parameters [25]. In addition, they can amplify evanescent waves and serve as a lens with subwavelength resolution [25]. Particularly, as a brand-new mechanism distincts from PT symmetry, CMs with a purely imaginary form [28] have been demonstrated to realize CPA and laser modes simultaneously in a comprehensive way beyond normal incidence [29, 30]. Moreover, based on CPA and laser modes in purely imaginary metamaterials (PIMs) [31], perfectly bidirectional negative refraction can be realized by utilizing such a pair of PIM slabs, which is a breakthrough to the unidirectional limit in PT symmetric systems. Therefore, these works [28-31] also provide alternative evidence that PT-symmetry is not necessary condition for achieving the coexistence of CPA and laser modes, which has been indicated in Ref. [32, 33].

Inspired by these results, in this work we introduce CMs to explore the CPA and laser modes in two-dimensional cylindrical structures. We find that in a CM cylinder, CPA and laser modes can be realized when the required material properties of CMs are satisfied. More interestingly, as a unique feature compared with other non-Hermitian optical systems, the required phase factors of CMs for obtaining CPA and laser modes are determined by the geometric size of the CM cylinder. Based on this property, we also find that CPA and laser modes can coexist in a CM cylinder with an extremely large size, where the excitation of CPA and laser modes depends on the angular momentum of incident wave. Therefore, based on the brand-new non-Hermitian systems, i.e., CMs, the simultaneous realization of CPA and laser modes, which is previously limited in the PT symmetric layered systems [11-13], is extended to cylindrical systems for the first time. Such a concept of obtaining CPA and laser modes simultaneously in cylindrical structures is significant and worthy of exploring in other non-Hermitian optical systems.

## 2. Theory and model for CPA and laser modes in CM cylinder

Firstly, let us consider a simple scenario shown in Fig. 1, where an infinite CM cylinder (green region) with its axis along $z$ direction is embedded in air. Its radius is $R_1$, and the EM parameters of CMs are given as $\varepsilon_1 = n_{CM} \exp(i\alpha)$, $\mu_1 = n_{CM} \exp(-i\alpha)$, where $n_{CM}$ and $\alpha \in (0, \pi)$ are the amplitude and phase factor of CMs respectively. Although the CMs are conventional materials from the view of refractive index, the loss and gain are included in CMs simultaneously. Hence, CPA and laser modes in principle might be obtained in the CM cylinder for incident waves with angular momenta. For example, if the incident wave illuminates the CM cylinder, and when the loss takes effect, CPA modes can be achieved without any scattering wave (see Fig. 1(a)). On the contrary, when the gain is dominant, laser modes can be obtained with intense outgoing wave (see Fig. 1(b)). To verify the above assumptions, we explore these results by analyzing the field distributions of the proposed structure in Fig. 1. Here the transverse magnetic (TM) with magnetic field in the $z$-direction is considered; for the transverse electric (TE) wave, the inverse results can be obtained due to the symmetry of Maxwell's equations as well as the time reversal symmetry of CM parameters. For TM wave, the total magnetic field outside the cylinder, is composed of two parts: the incident wave (with the Hankel functions of the second kind) and the scattered wave (with the Hankel functions of the first kind), i.e.,

$$\vec{H}_{total} = \hat{z} \sum_{m=-\infty}^{+\infty} \left[ H_m^{(2)}(k_0 r) \exp(im\theta) + S_m H_m^{(1)}(k_0 r) \exp(im\theta) \right], \quad (r \geq R_1) \tag{1}$$

where $k_0 = 2\pi/\lambda_0$ is wavevector in air, $\lambda_0$ is the vacuum wavelength, $m$ is the order of Hankel functions, and $S_m$ are the scattering coefficients. Inside the cylinder ($r<R_1$), the magnetic field is expressed as,

$$\vec{H}_{inside} = \hat{z}\sum_{-\infty}^{+\infty} a_m J_m(k_1 r)\exp(im\theta), \quad (0<r<R_1) \tag{2}$$

where $J_m(\cdot)$ is the Bessel functions, $a_m$ are coefficients to be determined, and $k_1 = k_0\sqrt{\varepsilon_1\mu_1} = n_{CM}k_0$ is a real wavevector in the cylinder. By matching the boundary conditions, we can get the scattering coefficient for each order,

$$S_m = \frac{k_1 J'_m(k_1 R_1) H_m^{(2)}(k_0 R_1) - \varepsilon_1 k_0 J_m(k_1 R_1) H'^{(2)}_m(k_0 R_1)}{\varepsilon_1 k_0 J_m(k_1 R_1) H'^{(1)}_m(k_0 R_1) - k_1 J'_m(k_1 R_1) H_m^{(1)}(k_0 R_1)}, \tag{3}$$

where $J'_m(\cdot)$ ($H'_m(\cdot)$) is the first-order derivative of the Bessel (Hankel) functions. Commonly $S_m=0$ indicates CPA modes, while $S_m \to \infty$ leads to laser modes. Based on Eq. (3), the conditions for obtaining CPA and laser modes can be further given as,

$$\begin{aligned} C_m^{(1)} &= \eta V_m, \quad \text{(laser)} \\ C_m^{(2)} &= \eta V_m, \quad \text{(CPA)} \end{aligned} \tag{4}$$

where $C_m^{(1)} = H_m^{(1)}(k_0 R_1)/H'^{(1)}_m(k_0 R_1)$ is the purely scattering information associated with laser effect, $C_m^{(2)} = H_m^{(2)}(k_0 R_1)/H'^{(2)}_m(k_0 R_1)$ is solely the portion of incident waves in vacuum corresponding to CPA effect, $V_m = J_m(n_{CM}k_0 R_1)/J'_m(n_{CM}k_0 R_1)$ is related to the cavity mode oscillation inside the CM cylinder, and $\eta = \varepsilon_{CM} k_0/k_1 = \exp(i\alpha)$ is the complex relative impedance between air and the CMs.

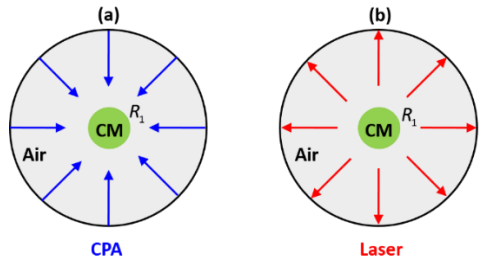

Figure1. The schematic diagrams of obtaining CPA modes (a) and laser modes (b) in a CM cylinder by considering the incident wave with angular momentum.

It is noted that both terms $C_m^{(1)}$ and $C_m^{(2)}$ are complex numbers, which are only determined by the cylinder size $R_1$ for a fixed working frequency. To illustrate these two quantities more clearly, we re-express them as $C_m^{(1)} = \xi_m^{(1)} \exp(i\phi_m^{(1)})$ and $C_m^{(2)} = \xi_m^{(2)} \exp(i\phi_m^{(2)})$, respectively. Figure 2 shows the relationship between $C_m^{(1)}$ ($C_m^{(2)}$) and the radius $R_1$. Obviously in Fig. 2(a), both $C_m^{(1)}$ and $C_m^{(2)}$ have the same amplitude, i.e., $\xi_m^{(1)} = \xi_m^{(2)} = \xi_m$. For their phases as shown in Fig. 2(b), $\phi_m^{(1)}$ is always negative, located in the region of $\phi_m^{(1)} \in [-\pi, -0.5\pi]$; $\phi_m^{(2)}$ is always positive, located in the region of $\phi_m^{(2)} \in [0.5\pi, \pi]$; meanwhile $\phi_m^{(1)} = -\phi_m^{(2)}$. Therefore for a fixed geometry size, $C_m^{(1)}$ and $C_m^{(2)}$ share the same amplitude, yet with opposite phase. In this way, Eq. (4) can be further written as,

$$\xi_m(k_0 R_1) \exp[i\phi_m^{(1)}] = V_m(n_{CM} k_0 R_1) \exp(i\alpha), \quad \text{(laser)}$$
$$\xi_m(k_0 R_1) \exp[i\phi_m^{(2)}] = V_m(n_{CM} k_0 R_1) \exp(i\alpha), \quad \text{(CPA)}. \quad (5)$$

Here the term $V_m$ is a real number ranging from minus infinity to plus infinity, that is $V_m \in (-\infty, \infty)$. With these, Eq. (5) can be met by adjusting the properties of CMs, i.e., $\alpha$ and $n_{CM}$. After simple analysis, there are four situations for obtaining CPA and laser modes, with the corresponding phases defined as $\alpha_C$ and $\alpha_L$ respectively. (*i*) $\alpha_L = \phi_m^{(1)} < 0$ and $V_m = \xi_m$ for laser; $\alpha_C = \phi_m^{(2)} > 0$ and $V_m = \xi_m$ for CPA. In this case, $\alpha_L = -\alpha_C$, and both laser and CPA effect share the same index $n_{CM}$. (*ii*) $\alpha_L = \pi + \phi_m^{(1)} > 0$ and $V_m = -\xi_m$ for laser; $\alpha_C = \phi_m^{(2)} > 0$ and $V_m = \xi_m$ for CPA. In this case, $\alpha_L + \alpha_C = \pi$, and both laser and CPA happen at different index $n_{CM}$. (*iii*) $\alpha_L = \phi_m^{(1)} < 0$ and $V_m = \xi_m$ for laser; $\alpha_C = \pi + \phi_m^{(2)} = \phi_m^{(2)} - \pi < 0$ and $V_m = -\xi_m$ for CPA. Accordingly, $\alpha_L + \alpha_C = -\pi$ and both laser and CPA effect occur at different $n_{CM}$. (*iv*) $\alpha_L = \pi + \phi_m^{(1)} > 0$ and $V_m = -\xi_m$ for laser; $\alpha_C = \pi + \phi_m^{(2)} = \phi_m^{(2)} - \pi < 0$ and $V_m = -\xi_m$ for CPA. Accordingly, $\alpha_L = -\alpha_C$ and both laser and CPA effect occur at the same $n_{CM}$. Therefore, independent of the conditions, the phase factors of CMs for obtaining CPA and laser modes are only determined by the geometric size of the CM cylinder.

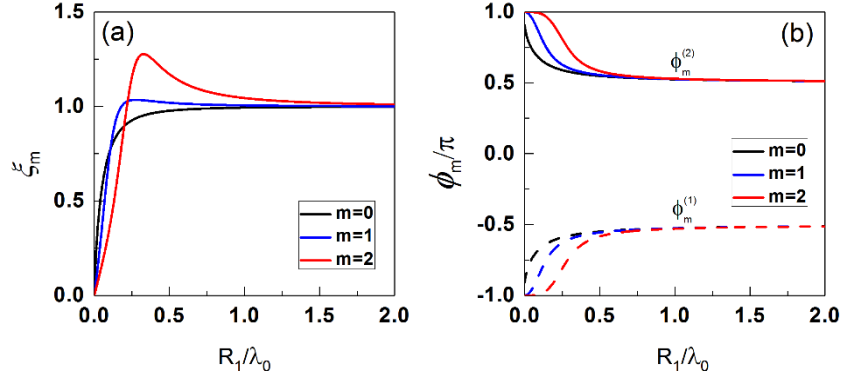

Figure 2. (a) The amplitude $\xi_m$ of $C_m^{(1)}$ and $C_m^{(2)}$ vs the radius $R_1$ of the CM cylinder. The black, blue and red curves are for the corresponding modes for $m$=0, 1 and 2 respectively. (b) The phase $\phi_m$ of $C_m^{(1)}$ and $C_m^{(2)}$ vs the radius $R_1$ of the CM cylinder. The black, blue and red curves are the corresponding modes for $m$=0, 1 and 2, respectively, yet with $\phi_m^{(1)}$ in dashed and $\phi_m^{(2)}$ in solid.

## 3. CPA and laser modes under different conditions

We consider CMs of the form $\varepsilon_1 = n_{CM}\exp(i\alpha)$ and $\mu_1 = n_{CM}\exp(-i\alpha)$ with $\alpha \in [0, \pi]$. In this case only the second situation (i.e., $\alpha_L = \pi + \phi_m^{(1)} > 0$ and $\alpha_C = \phi_m^{(2)} > 0$, with $\alpha_L + \alpha_C = \pi$) is satisfied for the following explorations. To verify the above analysis, Fig. 3 shows the relationship between $|S_m|$ and both factors: $n_{CM}$ and $\alpha$ of CMs in a logarithmic scale, with $R_1 = 0.25\lambda_0$ set as an example. The corresponding results for the coherent incident wave with angular momentum of $m$=0, $m$=1 and $m$=2 are respectively displayed in Figs. 3(a)-3(c). In each case, for instance $m$=0, the vanishing $S_m$ are marked by square frames; the high amplitude $S_m$ are denoted by circle frames. These vanishing and intense values are related to CPA and laser modes respectively. Clearly, all CPA modes happen at the same phase with $\alpha_C \in [0.5\pi, \pi]$, but at different $n_{CM}$. For all laser modes, they also occur at the same phase with $\alpha_L \in [0, 0.5\pi]$, but at different $n_{CM}$. Meanwhile the condition $\alpha_L + \alpha_C = \pi$ is satisfied. More specifically, for the $m$=0 case shown in Fig. 3(a), $C_0^{(1)} \approx 0.926 e^{-i0.59\pi}$ and $C_0^{(2)} \approx 0.926 e^{i0.59\pi}$ for $R_1 = 0.25\lambda_0$, leading to $|V_0| \approx 0.926$ for realizing CPA and laser modes. Based on Eq. (5), we can get $\alpha_C = 0.59\pi$ for all the CPA modes and $\alpha_L = 0.41\pi$ for all the laser modes.

Similar results can be observed for other cases of *m*=1 and *m*=2 (see Fig. 3(b) and Fig. 3(c)). Indeed, the required phase factors for CPA and laser modes are determined by the geometric size of the CM cylinders. Therefore, for any coherent incident waves, a CM cylinder with designed geometry size and the proper EM parameters can perfectly absorb all incident waves without any scattering, or can give rise to the laser phenomenon with intense outgoing waves.

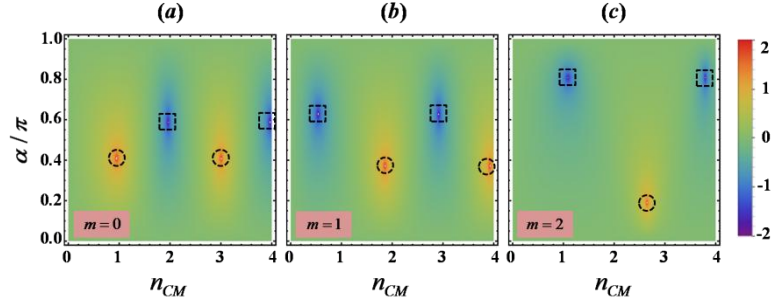

Figure 3. The scattering coefficients $|S_m|$ vs the amplitude $n_{CM}$, the phase factor $\alpha$ of CMs in a logarithmic scale. (a) *m*=0 ; (b) *m*=1; (c) *m*=2. In plots, the white regions mean that the values are beyond color bar. In all calculations $R_1=0.25\lambda_0$.

To verify the above analytical findings, numerical simulations are carried out by using COMSOL Multiphysics based on finite element method, with required parameters for CPA and laser modes indicated in Fig. 3. Figures 4(a)-4(c) show simulated magnetic field patterns, corresponding to three cases *m*=0, 1 and 2, respectively. In each case, for example *m*=0, the upper plot is the case of CPA mode and the lower one is the case of laser mode. From the upper field pattern in Fig. 4(a), the incident wave is totally absorbed in the CM cylinder without any scattering. Such a result is also confirmed consistently by the energy flows marked by the black arrows. In contrast, from the lower row of field pattern in Fig. 4(a), the laser mode is excited for the outgoing waves with high amplitude (see the color bar). Such a laser mode is also revealed by the outgoing energy flows. From Fig. 4(b) and Fig. 4(c), we can observe similar results for CPA and laser modes for *m*=1 and *m*=2. Therefore, CPA and laser modes can be realized in the CM cylinder when the required parameters of CMs are satisfied. It needs to mention that from the above analysis both CPA and laser effects actually happen in different CMs, i.e., one CM medium supports CPA, while another one is for laser.

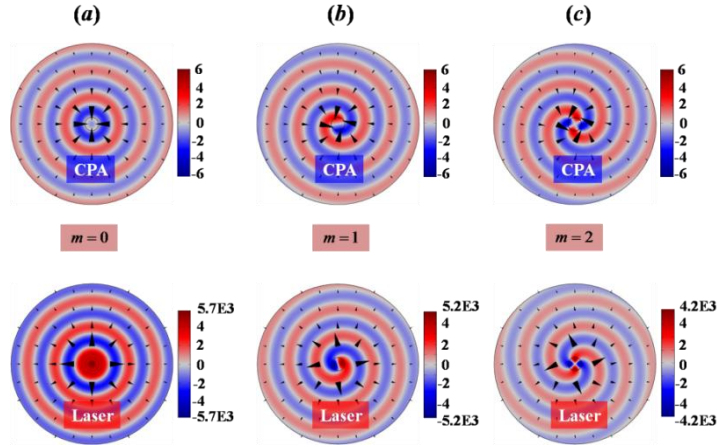

Figure 4. The simulated magnetic field patterns of CPA mode (upper plots) and laser mode (lower plots) for the incident wave with angular momentum. (a) *m*=0. For CPA, $n_{CM} \approx 1.96$ and $\alpha_C = 0.59\pi$ ; for laser $n_{CM} \approx 0.95$ and $\alpha_L \approx 0.41\pi$. (b) *m*=1. For CPA, $n_{CM} \approx 0.534$ and $\alpha_C \approx 0.63\pi$ ; for laser, $n_{CM} \approx 1.857$

and $\alpha_L \approx 0.37\pi$. (c) $m=2$. For CPA, $n_{CM} \approx 1.09$ and $\alpha_C \approx 0.81\pi$; for laser, $n_{CM} \approx 2.648$ and $\alpha_L \approx 0.19\pi$. Here $R_1 = 0.25\lambda_0$.

## 4. The coexistence of CPA and laser modes

Now we will discuss a special case in which a CM medium can support CPA and laser simultaneously. As shown in Fig. 2, $R_1/\lambda_0 \to \infty$ leads to $\xi_m \to 1$, $\phi_m^{(1)} \to -0.5\pi$ and $\phi_m^{(2)} \to 0.5\pi$. Because the radius of CM cylinder is much larger than the working wavelength, it approaches the geometric optics limit. In this case, when the required phases are applied in the second case (i.e., $\alpha_L + \alpha_C = \pi$), we have $\alpha_L = \alpha_C = 0.5\pi$. In this way, the studied CM becomes $\varepsilon_1 = in_{CM}$ and $\mu_1 = -in_{CM}$, which is a special CMs, i.e., purely imaginary metamaterial (PIM) [28, 31]. To match the amplitude $\xi_m = 1$ for CPA and laser modes, the following conditions should be met, i.e., $V_m = -1$ for laser mode and $V_m = 1$ for CPA mode. It is noted that for a fixed working frequency, the mode states $V_m = J_m(\rho)/J'_m(\rho)$ ($\rho = n_{CM} k_0 R_1$) is a function of the index $n_{CM}$ of the CMs and the cylindrical geometry, which reduces to the following formula for $R_1/\lambda_0 \to \infty$,

$$V_m(\rho) = \begin{cases} \tan(\rho - \pi/4), & (m = even) \\ -\cot(\rho - \pi/4), & (m = odd). \end{cases} \tag{6}$$

These formulas tell us that the term $V_m$, related to the cavity mode oscillations, possesses identical properties for the incident waves with the same parity, i.e., they share the same resonance conditions for all even/odd modes. The result of $V_m$ vs $n_{CM}$ is shown in Fig. 5(a), where we set $R_1 = 100\lambda_0$ to approach the condition of $R_1/\lambda_0 \to \infty$. For even orders, for example $m=0$ and 2, both curves are coincident; for odd orders (for example $m=1$ and 3), both curves are also coincided. Concerning CPA and laser modes, Eq. (6) indicates two different cases. One is that when $V_{m=odd}(\rho) = 1$ is obtained by adjusting the index $n_{CM}$, the CPA effect happens for any incident wave with odd order $m$; meanwhile, $V_{m=even}(\rho) = -1$ is spontaneously satisfied, implying that the laser effect takes place for any incident wave with even order (e.g., see the case of $n_{CM} = 1.0975$ in Fig. 5(a)). Another case is that when $V_{m=even}(\rho) = 1$ is met, the CPA effect happens for any incident wave with even order $m$; accordingly, $V_{m=odd}(\rho) = -1$ is spontaneously satisfied, resulting in laser modes for any incident wave with the odd order (e.g., see the case of $n_{CM}=1.1$ in Fig. 5(a)). To illustrate this point clearly, we display the relationship between $|S_m|$ and $n_{CM}$ shown in Fig. 5(b). For the even modes, e.g., $m=0$ (the black solid curve) and $m=2$ (the red dashed curve), CPA modes will happen at $n_{CM} \cong 1.1$; while for odd modes e.g., $m=1$ (the blue solid curve) and $m=3$ (the green dashed curve), laser modes also take place at $n_{CM} \cong 1.1$. Likewise, for the case of $n_{CM} \cong 1.0975$, CPA modes can happen for odd modes, while laser modes can occur for even modes. Therefore, CPA and laser modes can function simultaneously in such an extreme case, where the excitations of CPA and laser modes depend on the parity of the angular momenta of the incident coherent waves. To the best of our knowledge, this is a novel feature that has not been reported in other media or optical systems including PT symmetric configurations.

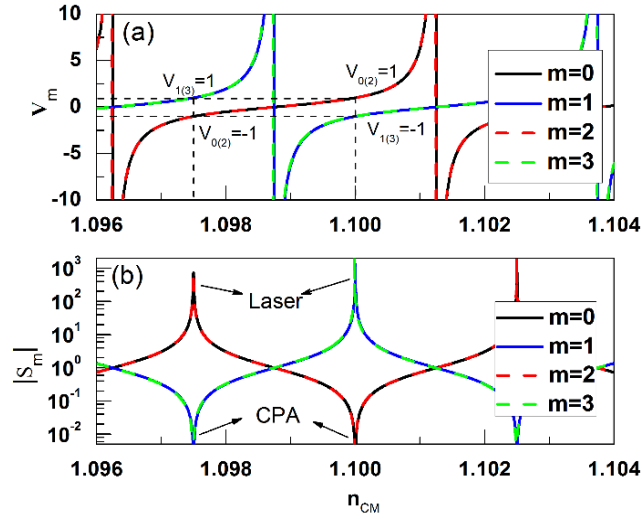

Figure 5. (a) The mode state $V_m$ vs the amplitude $n_{CM}$ of CMs. The black solid, blue solid, red dashed, and green dashed curves are the corresponding modes for $m = 0, 1, 2$ and $3$, respectively. (b) The scattering coefficient $|S_m|$ vs the amplitude $n_{CM}$ of CMs. The black solid, blue solid, red dashed, and green dashed curves are the corresponding modes for $m=0, 1, 2,$ and $3$, respectively. In all analysis, $\alpha=0.5\pi$ and $R_1=100\lambda_0$.

## 5. Conclusion

In conclusion, we have shown that by applying the concept of CMs into a two-dimensional cylindrical structure, CPA and laser modes can be realized by employing incident waves with specific angular momenta. In particular, a unique feature different from other non-Hermitian optical systems is that the required phase factors of CMs for obtaining CPA and laser modes are dependent on the geometric radius of the CM cylinder. We also analytically found that CPA and laser modes can coexist in a PIM cylinder within the geometric optics limit (i.e., larger geometric size). It seems that the coexistence of CPA and laser modes in a single cylinder with extreme profiles is not implemented with ease. As a compromise, such a problem can be well solved by using core-shell cylindrical structures.

As loss and gain media are simultaneously included in uniform media, CMs remain challenging to realize in practice via current metamaterial technique, but are still possible. Recently, two schemes [29,

30] to realize materials with properties similar to CMs have been proposed in theory. For instance, by utilizing PT-symmetric metasurface configuration [29], where the loss and gain are individually distributed in different layers, CMs can be effectively mimicked. Moreover, on basis of the effective medium theory, a partial of CMs can be theoretically designed by employing photonic crystal of core-shell rods with loss and gain media [30]. Another possible solution is inserting a series of subwavelength 3D particles or 2D cylinders made of gain dielectric media in a lossy background medium. For instance, in a 2D composite structure composed of a lossy epsilon-near-zero (ENZ) medium with a gain-dielectric cylinder, an effective permeability with negative imaginary part (gain) [34] can be designed via the resonance of cavity modes. Therefore, by adjusting parameters of the geometry and materials, one might even realize a CM with near zero index.

In this work CMs were investigated in a narrow band of frequencies, but in fact they should have strong dispersion. Based on the broadband CPA applications in previous works [35, 36], similar feature for CMs (e.g., broadband CPA-laser modes) is also anticipated. In principle, similar analysis could be implemented in 3D spherical structures made of CMs by replacing Bessel and Hankel functions with their spherical analogs [37]. Therefore, more fantastic optical properties of CMs in 2D or 3D structure remain open to explorations, such as multi-channel CPA, multi-channel laser, and even multi-channel CPA-laser in a multi-layered cylindrical/spherical CM structure, which has potential applications in particle probing, optical commutations and ultrasensitive detection or sensing.

**Acknowledgement**

This work was supported by the National Natural Science Foundation of China (grant No. 11604229), the Natural Science Foundation of Jiangsu Province (grant no. BK20171206), the Postdoctoral Science Foundation of China (grant no. 2015M580456), and the Fundamental Research Funds for the Central Universities (Grant No. 20720170015). Y. Xu would like to thank the support from the Collaborative Innovation Center of Suzhou Nano Science and Technology at Soochow University.